\documentclass[10pt,conference ]{IEEEtran}
\usepackage{amsfonts}
\usepackage{times}
\usepackage{graphicx}
\usepackage{latexsym}
\usepackage{dsfont}
\usepackage{amssymb}
\usepackage{amsmath}
\usepackage{cite}
\usepackage{verbatim}
\usepackage{subfigure}

\def\bb0{{\mathbb{0}}}

\def\bb{{\mathbf{b}}}

\def\bg{{\mathbf{g}}}

\def\b0{{\mathbf{0}}}

\def\bB{{\mathbf{B}}}

\def\bX{{\mathbf{X}}}

\def\cD{\mathcal{D}}

\def\cG{\mathcal{G}}

\def\cS{\mathcal{S}}

\def\sf0{{\mathsf{0}}}

\usepackage{epstopdf}
\usepackage{enumerate}
\usepackage{algorithmicx}
\usepackage{algorithm}
\usepackage{amsmath}
\usepackage[noend]{algpseudocode}
\usepackage{float}
\usepackage{hyperref}
\usepackage{color}
\usepackage{makeidx}
\usepackage{bbm}
\usepackage{graphicx}
\usepackage{lipsum}
\usepackage{subfigure}
\usepackage{tablefootnote}
\usepackage{multirow}
\usepackage{multicol}
\usepackage{balance}
\usepackage{booktabs}
\usepackage{soul}
\usepackage[normalem]{ulem}
\usepackage{xcolor}

\newcommand{\comm}[1]{}

\begin{document}

\title{Pixel-Level GPS Localization and Denoising using Computer Vision and 6G Communication Beams}
\author{Gouranga Charan, Tawfik Osman, and Ahmed Alkhateeb\\Arizona State University - Emails: \{gcharan, tmosman, alkhateeb\}@asu.edu }

\maketitle

\begin{abstract}
Accurate localization is crucial for various applications, including autonomous vehicles and next-generation wireless networks. However, the reliability and precision of Global Navigation Satellite Systems (GNSS), such as the Global Positioning System (GPS), are compromised by multi-path errors and non-line-of-sight scenarios. This paper presents a novel approach to enhance GPS accuracy by combining visual data from RGB cameras with wireless signals captured at millimeter-wave (mmWave) and sub-terahertz (sub-THz) basestations. We propose a sensing-aided framework for (i) site-specific GPS data characterization and (ii) GPS position de-noising that utilizes multi-modal visual and wireless information. Our approach is validated in a realistic Vehicle-to-Infrastructure (V2I) scenario using a comprehensive real-world dataset, demonstrating a substantial reduction in localization error to sub-meter levels. This method represents a significant advancement in achieving precise localization, particularly beneficial for high-mobility applications in 5G and beyond networks.
\end{abstract}


\section{Introduction} \label{sec:intro}
Accurate location information is pivotal for a wide range of current and future applications, including autonomous vehicles, emergency services, and high-frequency 5G and beyond networks. The wide-scale availability of efficient localization using Global Navigation Satellite Systems (GNSS) and, in particular, Global Positioning Systems (GPS) has led to the large-scale adoption of GPS in various real-world systems. \textbf{However, the challenges of reliability and accuracy pose significant hurdles.} The publicly-available GNSS systems often suffer from large-scale errors ranging between 1-5 meters, primarily due to multi-path, non-line-of-sight (NLOS) scenarios, clock synchronization mismatches, and inherent device variations. The multi-path error results from receiving both the reflected and the line-of-sight (LOS) signals, while NLOS errors occur from signal reflections without a direct path. These inaccuracies significantly degrade the performance of systems that rely on precise location data, rendering them unsuitable for critical applications.

In response to these challenges, the past decade has seen a surge in efforts to develop more accurate location solutions for both indoor and outdoor settings \cite{Tranquilla1994, Jiang2014, Han2013, YIU2017235}. These efforts have generally focused on two main strategies: (i) hardware optimization to enhance existing GNSS systems \cite{Tranquilla1994, Jiang2014} and (ii) Radio Frequency (RF)-based solutions for improved localization \cite{Han2013, YIU2017235}. The hardware-based solution focuses on reducing positional error by optimizing the design of the transmitters and receivers and implementing advanced signal processing techniques. Innovations such as choke-ring  \cite{Tranquilla1994} and dual-polarization antennas \cite{Jiang2014} have proven effective in mitigating multi-path errors in GPS measurements. Although effective, their utility is limited in high-mobility environments. Further, their adoption is constricted by the high cost and the large size. 

The RF-based solutions, employing technologies such as ultra-wideband (UWB), WiFi, Radio Frequency Identification (RFID), and cellular networks (LTE and 5G), utilize fingerprint-matching techniques for accurate localization. These techniques depend on databases of signal "fingerprints" at specific locations. In general, the Received Signal Strength Indicator (RSSI) \cite{YIU2017235}, or Channel state information (CSI)-based fingerprint matching techniques, are widely adopted for accurate localization, especially for indoor scenarios. One of the significant challenges associated with these approaches is the need for pre-built RSSI or CSI maps. Any change in the environment deteriorates the model's performance, thereby impacting the reliability of such approaches. Furthermore, the RSSI and CSI-based approaches are inherently indoor-only solutions with minimal localization capability for dynamic outdoor locations.

\begin{figure}[t]
	\centering
	\includegraphics[width=0.9\columnwidth]{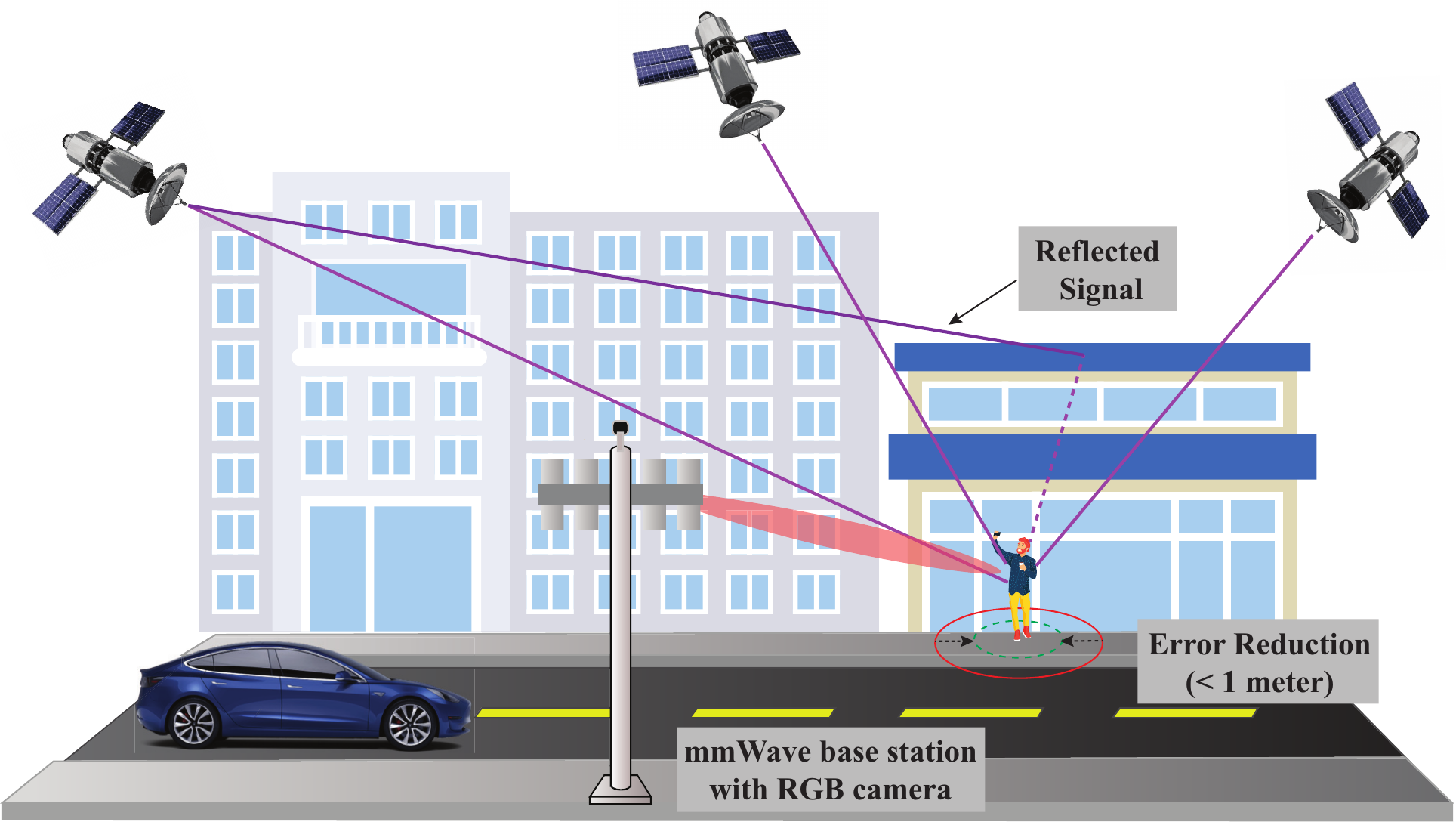}
	\caption{An illustration of the challenges associated with the GPS positioning resulting in positional errors of up to 10 meters in the real-world. }
	\label{fig:key_idea}
\end{figure}
Beyond $5$G  millimeter-wave (mmWave) and sub-terahertz (sub-THz) communication systems are envisioned to be equipped with an array of sensors such as RGB cameras \cite{charan2023camera}, radar \cite{demirhan2023integrated}, and LiDAR \cite{jiang2022lidar} to enable joint communication and sensing applications. This technological advancement opens up new possibilities for mitigating GPS measurement errors by utilizing additional sensing data gathered at basestations. Among these modalities, visual data from RGB cameras provides spatial and contextual information that, when integrated with GNSS, can significantly improve localization accuracy. Visual data offers a detailed environmental view unaffected by the common GNSS system errors such as multi-path and NLOS issues. Moreover, advancements in computer vision and machine learning now allow for the real-time extraction and processing of visual data. Motivated by this potential, our paper investigates the use of visual data in combination with wireless signals to enhance the accuracy of real-world GPS measurements significantly.

This paper proposes to mitigate the errors associated with practical GPS measurements by leveraging both wireless and visual data captured at the mmWave/sub-THz basestation. The main contributions of the work can be summarized as follows:
\begin{itemize}
	\item Formulating the sensing-aided position de-noising problem considering practical GPS measurements and communication models.
	\item Developing a grid-based approach using visual and wireless communication data to perform site-specific characterization of real-world GPS measurements. 
	\item Evaluating the performance of the proposed solution in a realistic Vehicle-to-Infrastructure (V2I) scenario based on our large-scale real-world dataset, DeepSense 6G \cite{DeepSense} that consists of a co-existing multi-modal sensing and communication dataset.
\end{itemize}

Based on the adopted real-world dataset, the developed solution can help reduce the error to a sub-meter level. This highlights the capability of the proposed sensing-aided position de-noising approach.

\section{Vision-Wireless Position De-noising:\\ System Model and Problem Formulation}

Building on the premise that integrating visual and wireless data can significantly enhance localization accuracy, we delve into the specifics of the system model and problem formulation for GPS position de-noising in this section.

\subsection{System Model} \label{sec:sys_ch_mod}

This paper adopts the system model illustrated in Fig.~\ref{fig:key_idea}, where a basestation equipped with an RGB camera is serving a mobile vehicle. The basestation is equipped with a uniform linear array (ULA) with $M$ elements and an RGB camera. This basestation serves a mobile user that is equipped with, for simplicity, a single antenna and a GPS receiver. The basestation adopts a predefined local beamforming codebook $\boldsymbol{\mathcal F}=\{\mathbf f_q\}_{q=1}^{Q}$, where $\mathbf{f}_q \in \mathbb C^{M\times 1}$ and $Q$ is the total number of beamforming vectors. The communication system in this work adopts OFDM with a cyclic prefix of length $D$ and $K$ sub-carriers. The downlink received signal at the mobile unit is given by:
\begin{equation}\label{eq:sys_mod}
	y_{k} = \mathbf h_{k}^T \mathbf f^{\star} x + v_k,
\end{equation}
where $y_{k}\in \mathbb C$ is the received signal at the $k$th sub-carrier, $\mathbf f^{\star}\in \mathcal F$ is the optimal beamforming vector, $\mathbf h_{k} \in \mathbb C^{M\times 1}$ is the channel between the BS and the mobile unit at the $k$th sub-carrier, $x\in \mathbb C$ is a transmitted complex symbol that satisfies the following constraint $\mathbb E\left[ |x|^2 \right] = P$, where $P$ is a power budge per symbol, and finally $v_k$ is a noise sample drawn from a complex Gaussian distribution $\mathcal N_\mathbb C(0,\sigma^2)$.

\begin{figure}[t]
	\centering
	\includegraphics[width=0.8\columnwidth]{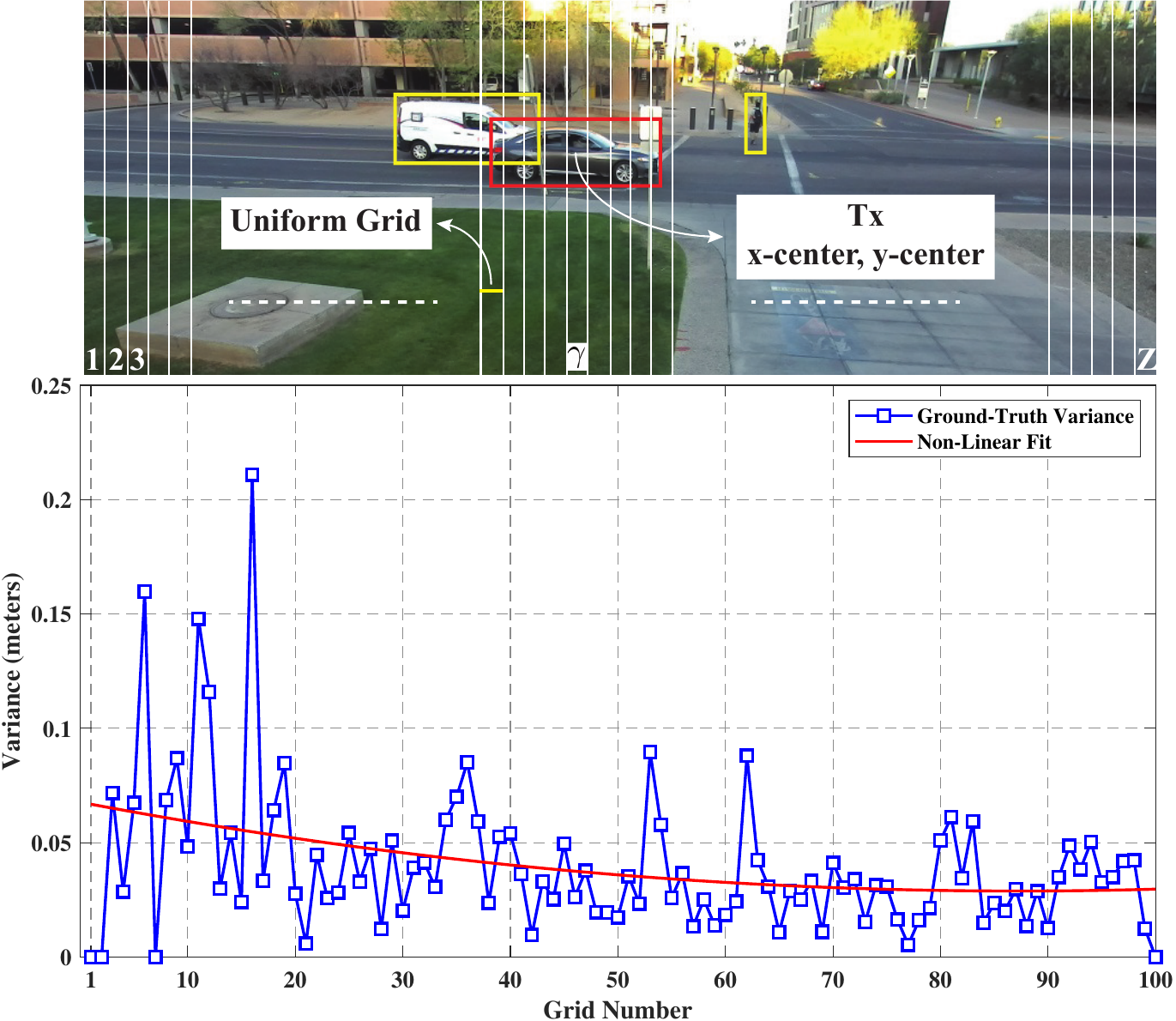}
	\caption{This figure illustrates the grid-based clustering approach adopted in this work, where the wireless environment is divided into $Z$ uniform grids.} 
	\label{fig:grid_var_gt}
\end{figure}

\subsection{Problem Formulation} \label{sec:prob_form}
GPS accuracy is influenced by several factors, including the need for signals from at least $4$ satellites \cite{GPS_accuracy} and errors introduced by atmospheric conditions, receiver noise, and multi-path effects. Our goal is to enhance GPS data precision to a sub-meter level, leveraging (i) RGB imagery from the basestation's camera, (ii) optimal wireless beam indices, and (iii) the user's initial noisy GPS data. For that, we define $\widetilde \bg \in \mathbb R^2$ as the two-dimensional noisy position vector (carrying latitude and longitude information). Further, we define $\bX \in \mathbb{R}^{W \times H \times C}$ as the RGB image of the scene captured at the basestation, where $W$, $H$, and $C$ are the width, height, and the number of color channels for the image. We formulate the task of position de-noising from a machine learning perspective as a regression problem. The problem can be defined as learning a mapping $f_{\theta} : \cS \rightarrow \widetilde \cG$ from an input space $S$ to a continuous output space $\widetilde \cG$, for a set of example pairs $\{( s_r,\widetilde \bg_r)\}_r \subset \cS \times \widetilde \cG$, where $\cS = \left (\bX_r, \mathbf f^{\star}_r \right )$ and $r \in [1, R]$ represent each data sample in dataset $\cD$. The model is developed to learn the function $f_\theta \left (\cS \right )$ parameterized by $\theta$ (e.g., the weights of a deep neural network) by minimizing a loss function. The loss function $L(\theta)$ is defined as
\begin{equation}
	L(\theta) = \frac{1}{R}\sum_{r=1}^{R}  l \left (f_\theta \left ( s_r \right ), \ \widetilde\bg_r \right),
\end{equation}
where $R$ is the total number of data samples in dataset $\cD$. The model takes in the observed image-beam pair and predicts the de-noised position $\hat \bg \in \mathbb R^2$. The function $l$ measures the deviation of the prediction value from the corresponding ground-truth value. Next, we present the proposed grid-based measurement grouping in Section~\ref{sec:grid_analysis}.

\begin{figure*}[t]
	\centering
	\includegraphics[width=0.8\linewidth]{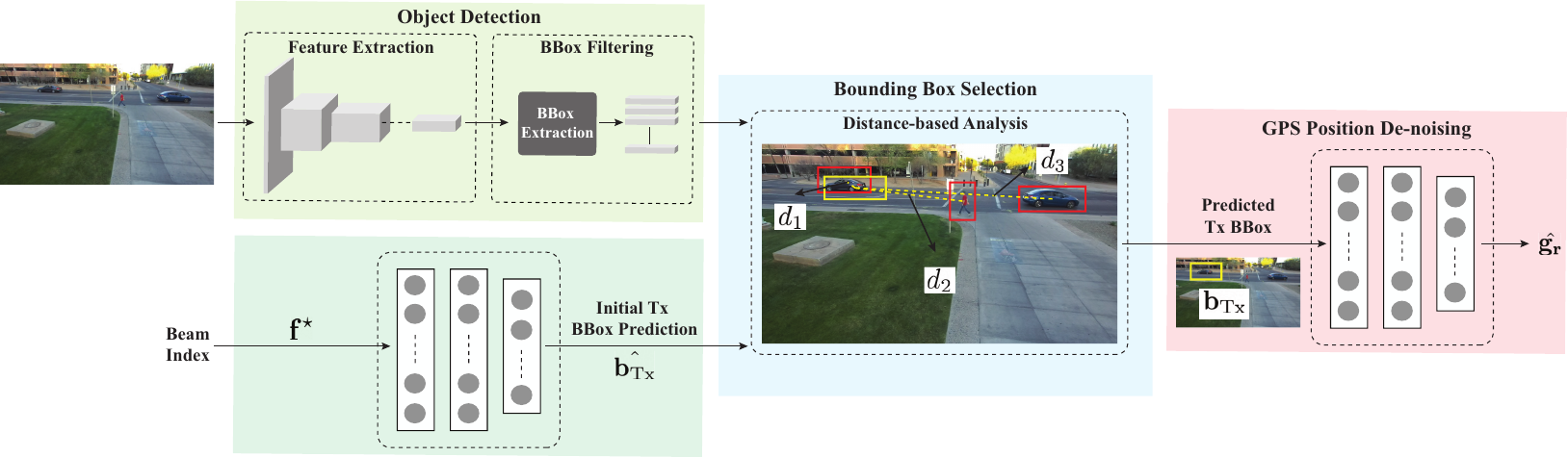}
	\caption{This figure presents the proposed 2-stage GPS position de-noising solution using the multi-modal visual and wireless data captured by the basestation. It highlights the two stages of the proposed solution, namely (i) transmitter identification and (ii) GPS position de-noising stage.}
	\label{fig:ml_model}
\end{figure*}

\section{Vision-Aided Measurement Grouping and Error Analysis}\label{sec:grid_analysis}
The inherent noise and errors in GPS receivers complicate the accurate analysis and characterization of GPS data. To address these challenges, we introduce a vision-aided, grid-based measurement grouping and error analysis method. Grid-based strategies have historically facilitated the clustering of noisy, high-dimensional data. Our objective is to segment the data into clusters where each cluster's positional data correspond to a singular real-world location.

\textbf{Vision-Aided Grouping:} The RGB camera present at the basestation captures images of the environment in the field of view of the basestation. The first step in the proposed algorithm is dividing the image into $Z$ equal vertical grids. Each grid, therefore, has a width of $1/Z$ units, considering the total width of the image to be $1$. The critical question is, \textit{how do we accurately map each GPS position to a specific grid?} This proposed data grouping/mapping leverages advances in computer vision, specifically object detection techniques, to identify and extract normalized center coordinates (x-center, y-center) of objects within the scene. The detailed approach of extracting the center coordinates is presented in Section~\ref{sec:ml_soln}. The final step is assigning a grid value based on the x-center coordinate of the object of interest. For example, an object is assigned the grid $\gamma$ if 
\begin{equation}
	\frac{\gamma}{Z} \leq \text{x-center} < \frac{{\gamma + 1}}{Z},
\end{equation}
where $\gamma \leq {Z-1}$. This step effectively associates each data sample with a grid, setting the stage for error characterization.

\textbf{Error Characterization:} Once all the data samples are assigned a particular grid, the error associated with a particular GPS data can be characterized. For this, we propose a two-step approach: (i) First, for a particular grid $\gamma$, we compute the mean latitude and longitude values $\{lat^{\gamma}_m, long^{\gamma}_m\}$ from the associated data points  (ii) second, we compute the average displacement $d_{\gamma}$ between the mean and all the data samples associated with the particular grid $\gamma$
\begin{equation}
	d_{\gamma} = \frac{1}{N}\sum_{n=1}^{N} \textbf{hav} \left (\{lat^{\gamma}_m, long^{\gamma}_m\} - \{lat^{\gamma}_p, long^{\gamma}_p\} \right ),
\end{equation}
where $\text{hav}$ is the Haversine distance (equation \ref{eq:hav}) between two points and $N$ is the total number of data points in grid $\gamma$. The average displacement between the mean and the data points per grid can be considered as the displacement error associated with the GPS data. In order to perform the error analysis, we study the distribution of the average displacement across all the grids. Fig.~\ref{fig:grid_var_gt} shows the grid-based clustering of a visual scene and the error variance in the ground-truth GPS data (Left to Right movement of the transmitter). It is observed from the variance plot that the error across the entire dataset is not uniform. For example, some locations (specifically on the left side) have a higher variance in position data than the other locations. The insights gained from this analysis guide our multi-modal solution for GPS position de-noising, presented in the following section.

\section{Proposed Solution: Multi-Modal Vision-Wireless Based Position De-Noising} \label{sec:prop_sol}
In this section, we present an in-depth overview of the proposed position de-noising solution. First, we present the key idea in Section~\ref{sec:key_idea} and then explain the details of our proposed solution in Section~\ref{sec:ml_soln}. 

\subsection{Vision-Aided 6G Positioning - Key Idea} \label{sec:key_idea}

The high error margin of $\approx 5$ meters in GPS positional data poses a significant obstacle to enabling technologies like smart cities, autonomous vehicles, and enhanced road safety. This paper introduces a novel strategy to minimize this error to a sub-meter level, addressing the challenge posed by various external factors on positional accuracy. Diverging from conventional RSSI or CSI-based approaches, our solution employs machine learning with mmWave/THz wireless and visual data from a camera-equipped 6G communication system. We begin by harnessing recent advancements in computer vision to detect objects within the basestation's field of view. Next, we use wireless data to distinguish the transmitter from all the objects detected in the image. This dual approach of object detection and transmitter identification forms the basis of our proposed vision-aided position de-noising system. Ideally, a precise GPS system should consistently report the same location for an object if it returns to the same spot. However, GPS data often shows considerable variation due to its inherent inaccuracies. In contrast, the location of objects identified through visual data—achieved by object detection and transmitter identification—remains consistent. This stability allows us to use the visually identified locations as reliable anchors and helps reduce the overall error in the position data (provided a large enough dataset to capture the overall distribution of the noisy GPS dataset is available).

\subsection{Vision-Aided 6G Positioning - Proposed Solution} \label{sec:ml_soln}

This work addresses the challenge of GPS position de-noising in environments with multiple potential transmitters by proposing a two-stage solution: (i) transmitter identification through the integration of visual and wireless data and (ii) subsequent GPS position de-noising. Our approach utilizes advancements in machine learning and computer vision, alongside mmWave/THz data, to accurately identify transmitting candidates and refine GPS coordinates in real wireless settings. Fig.~\ref{fig:ml_model} presents the architecture of the proposed solution.

\subsubsection{Transmitter Identification:} 
The first stage of the proposed multi-candidate GPS position de-noising solution is to identify the transmitting candidate in the scene. Utilizing advancements in computer vision and machine learning, we can detect different objects in the environment and extract the relative position of the objects in the image. The visual data, further, needs to be augmented with some other modality to aid the transmitter identification task. The wireless beam index is the preferred modality in this work because the beamforming vectors provide directional information that summarizes the dominant signal direction for well-calibrated antenna arrays. This directional information can be projected onto the image plane, resulting in the form of image sectoring. The details of the two-step solution are presented next.

\textbf{Object detection:}  In order to perform object detection, a pre-trained YOLOv3 \cite{yolo} object detector is adopted. The pre-trained YOLOv3 model is further fine-tuned to detect two classes of objects in the scene, namely, ``Tx (Transmitter)'' and ``Distractors''. The fine-tuned YOLOv3 model is next utilized to extract the normalized bounding box center coordinates $\bB \in \mathbb R^{V \times 2}$ of the detected objects, where $V$ is the number of relevant objects detected in the scene. 

\textbf{Bounding box selection:} The second step in this pipeline is identifying the probable transmitter in the scene by utilizing the extracted bounding box matrix $\bB$ and the optimal beam index $\mathbf f^{\star}$. For this, we learn a prediction function $f_{\theta1}\left ( \cS \right ) $ that estimates the bounding box centers of the transmitting candidate using the beam indices. A 2-layered feed-forward neural network is developed to learn this prediction function,
\begin{equation}
	\hat \bb_{\text{Tx}} = f_{\theta1}(\mathbf f^{\star}_r | \bX_r),
\end{equation} 
where $\hat \bb_{\text{Tx}} \in \mathbb R^{1 \times 2}$ is a vector with the initial estimate of the center of the transmitting candidate and $\theta1$ are the parameters of the neural network. The initial prediction of the neural network $\hat \bb_{\text{Tx}}$ is then utilized to select the transmitting candidate in the scene $\bb_{\text{Tx}}$. This is done by calculating the Euclidean distance between each row of matrix $\bB$ and the predicted initial estimate. The object with the least distance to $\hat \bb_{\text{Tx}}$ is selected as the transmitting candidate ($\bb_{\text{Tx}}$) in the scene. 

\begin{figure}[!t]
	\centering
	\includegraphics[width=0.8\linewidth]{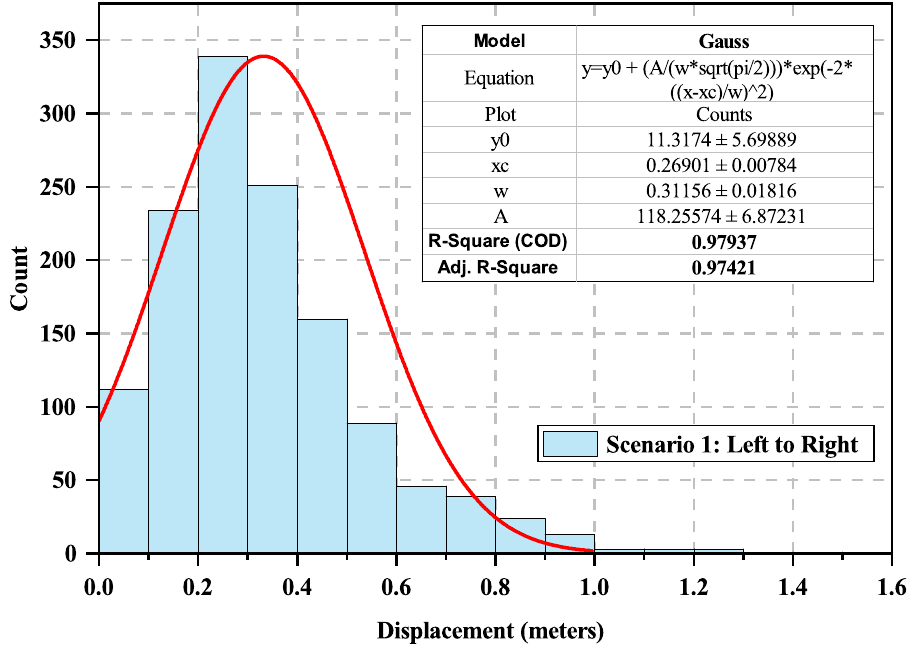}
	\caption{This figure plots the error distribution calculated for ground-truth GPS data for both left-to-right and right-to-left datasets. }
	\label{fig:gps_error}
	
\end{figure}

\begin{figure*}[!t]
	\centering
	\includegraphics[width=0.8\linewidth]{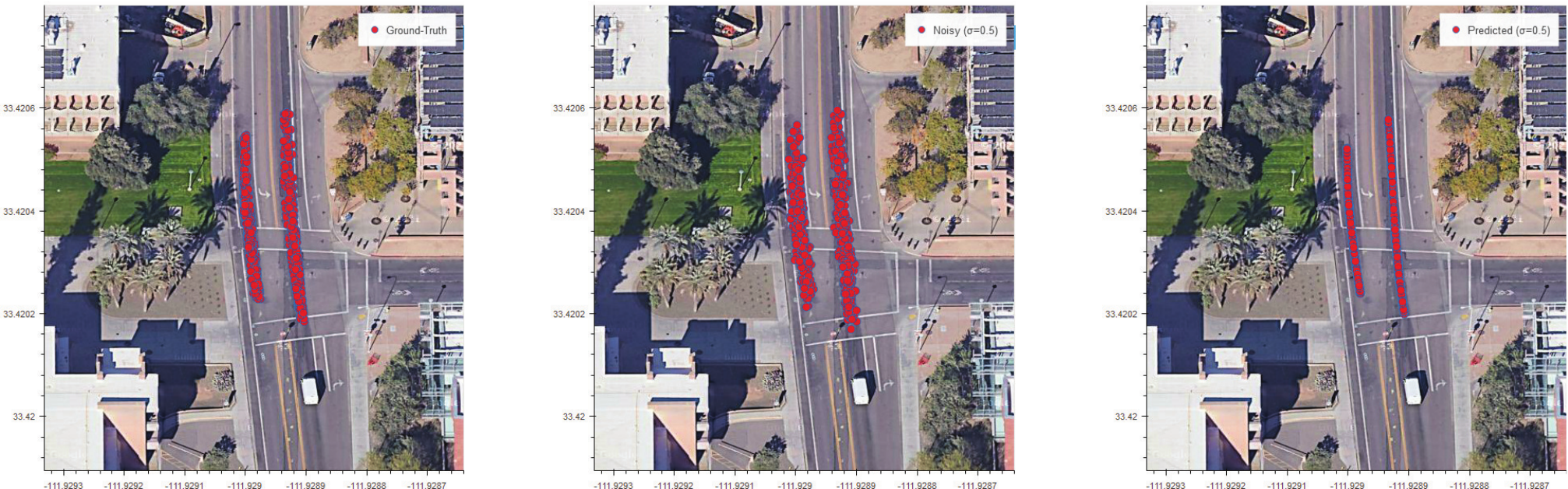}
	\caption{This figure overlays the ground-truth, noisy ($\sigma=0.5$) and de-noised position on the Google Earth satellite view for both the left-to-right and right-to-left datasets. The plot in the middle highlights the impact of adding noise to the GPS positions.  }
	\label{fig:gmap_plot}
	
\end{figure*}


\subsubsection{GPS Position De-noising}
Completing the first stage of the proposed solution helps identify the transmitting object in the scene. The next task is to predict the de-noised position of the transmitter. In this work, we explore two different approaches to address the position de-noising task, namely, (i) Lookup table (LUT)-based prediction and (ii) deep learning-based prediction. The two different approaches with varying computational complexity were selected to perform an in-depth comparative study. 

\textbf{LUT-based Prediction:} The LUT-based prediction strategy is grounded in the principle, as outlined in Section~\ref{sec:key_idea}, that GPS positions for a specific location tend to aggregate around their average values. This concept is reinforced by a grid-based method that organizes data samples into distinct, non-overlapping clusters, suggesting the average noisy GPS position is likely close to the actual ground-truth average. This insight forms the cornerstone of the LUT-based approach. Here, we use the x-coordinate of the final bounding box center ($\bb_{\text{Tx}}$) to estimate the transmitter's precise location. The process begins with constructing a lookup table from the training dataset, where each data sample, as detailed in Section~\ref{sec:grid_analysis}, is aligned with one of the $Z$ predefined grids. For each grid, we calculate the average latitude and longitude from its associated data samples, populating the LUT with these mean values alongside their corresponding grid indices. During the prediction phase, we identify the grid index for a given test sample based on its bounding box center ($\bb_{\text{Tx}}$) x-coordinate and retrieve the de-noised position from the pre-calculated mean values stored in the LUT.

\textbf{Deep learning-based prediction: } The deep learning-based proposed solution consists of a 2-layered feed-forward neural network that predicts the de-noised position of the transmitting candidate. The prediction task is posed as a regression problem, in which the input to the model is the center coordinates of the identified transmitting object, and the output is the de-noised position of the transmitter. The model is trained in a supervised fashion with the noisy GPS position as the labels. The goal here is to learn a function that can estimate the position of the transmitter using the predicted centers of the transmitting candidate ($\hat \bb_{\text{Tx}}$).

\section{Testbed Description and Development Dataset}\label{sec:dataset}
In order to evaluate the performance of the proposed solution, we adopt Scenario $1$ from the DeepSense $6$G dataset. It is the first large-scale real-world dataset comprising co-existing data modalities such as vision, LiDAR, Radar, mmWave wireless, and position. This section presents a brief overview of the scenario adopted from the DeepSense 6G dataset, followed by the analysis of the final development dataset utilized for the sensing-aided beam prediction study.

\subsection{DeepSense 6G: [Scenario 1]}  \label{sec:testbed}

The DeepSense testbed $1$ consisting of a stationary and a mobile unit is utilized for this data collection. The stationary unit \{unit1 (RX)\} is equipped with a standard-resolution RGB camera and mmWave Phased array. The stationary unit adopts a 16-element ($M = 16$) 60GHz-band phased array, and it receives the transmitted signal using an over-sampled codebook of $64$ pre-defined beams ($Q = 64$). In this data collection scenario, the mobile unit \{unit2 (TX)\} is a vehicle equipped with a mmWave transmitter, GPS receiver, and inertial measurement units (IMU). The transmitter consists of a quasi-omni antenna constantly transmitting (omnidirectional) at 60 GHz. For more information regarding the data collected testbed and setup, please refer to \cite{DeepSense}.

\subsection{Development Dataset} \label{sec:dev_dataset}

The initial data collected by the DeepSense-6G testbed is the raw data. The raw data undergoes multiple task-specific post-processing steps to generate the final development dataset. The first step is to divide the data into two different datasets based on the lane the transmitter traveled during the data collection process. In an ideal condition, the horizontal accuracy of the publicly available GPS position is in the range of $0.2 - 0.3$ meters. The data was divided to eliminate the shift in position induced by the passing lane, which has a width of $\approx 3.5$ meters. This would ensure a more accurate characterization of GPS error and help develop robust solutions to mitigate the same. The two datasets generated are (i) left-to-right dataset: This dataset comprises samples with the transmitter traveling from left to right of the basestation. (ii)Right-to-left dataset: The data collected when the transmitter traveled from the right to the left of the basestation. The second step in the post-processing pipeline is removing the outliers to ensure clean datasets. The final step is adding a zero-mean Gaussian noise to the GPS positions with different variance values. The different position variances are selected such that the resultant datasets have distance error variances of $\{0.1, 0.5, 1.0, 2.0, 3.0\}$ meters. The final left-to-right and right-to-left development datasets consist of $1353$ and $1086$ data samples. Both datasets are further divided into training and test sets with a $70-30\%$ split.

\section{Performance Evaluation}\label{sec:perf_eval}

This section evaluates the performance of the proposed multi-modal position de-noising solution. In the first sub-section, we discuss the adopted evaluation metrics. Next, we present the site-specific GPS error characterization analysis and evaluate the de-noising capability of the proposed solution.

\subsection{Evaluation Metrics}

In this section, we present the evaluation metric used to evaluate the performance of our proposed solution. We utilize the Haversine distance formula to calculate the deviation of the predicted position from the ground-truth position. The Haversine formula calculates the shortest distance between two points on a sphere using their latitudes and longitudes information. The Haversine distance between two positions is calculated as
\begin{equation}
	\textbf{hav}=2usin^{-1}\left (\sqrt{sin^{2} \left (\frac{\phi_2 - \phi_1}{2} \right ) + \Delta sin^{2}\left (\frac{\lambda_2 - \lambda_1}{2} \right )} \right ),
	\label{eq:hav}
\end{equation} 
where $ \Delta = cos\left (\phi_1 \right )cos\left (\phi_2 \right )$, $u$ is the radius of the earth(6371 km), $\phi_1, \phi_2$ are the latitudes of the two points and $\lambda_1, \lambda_2$ are the longitude of the two points, respectively.

\subsection{Numerical Results}\label{sec:results}
This section presents the site-specific GPS error characterization analysis and evaluates the position de-noising capability of the proposed solution. For this, we divide each visual scene into $100$ grids (i.e., $Z=100$) such that the width of each grid is $\approx 0.3$ meters.

\textbf{Can grid-based approach help site-specific GPS error Characterization?}
It is essential to characterize location-specific GPS errors to help develop more efficient solutions for accurate positioning systems. In this work, we characterize the GPS position error for both datasets consisting of the left-to-right and right-to-left movement of transmitting vehicles. To characterize the individual GPS error, we use the two-step approach presented in Section~\ref{sec:grid_analysis}. The two-step approach provides the average displacement of the ground-truth position with respect to the mean position per grid. Fig.~\ref{fig:gps_error} shows the distribution of the average displacement for both datasets. The peak in both datasets is observed to be around $0.3$ meters, further validating the choice of $Z$. We also modeled the data spread by assigning the best fit function for the entire range. The Gaussian distribution function is observed to have the best fit with an adjusted R-Square of $0.9742$ and $0.9738$ for left-to-right and right-to-left datasets. The observation is consistent with the previous studies \cite{GPS_error} done on characterizing GPS time-series data, which reported that the GPS error is a mixture of white noise and flicker noise.


\begin{figure}[!t]
	\centering
	\subfigure[Left-to-Right Dataset]{\centering \includegraphics[width=0.9\columnwidth]{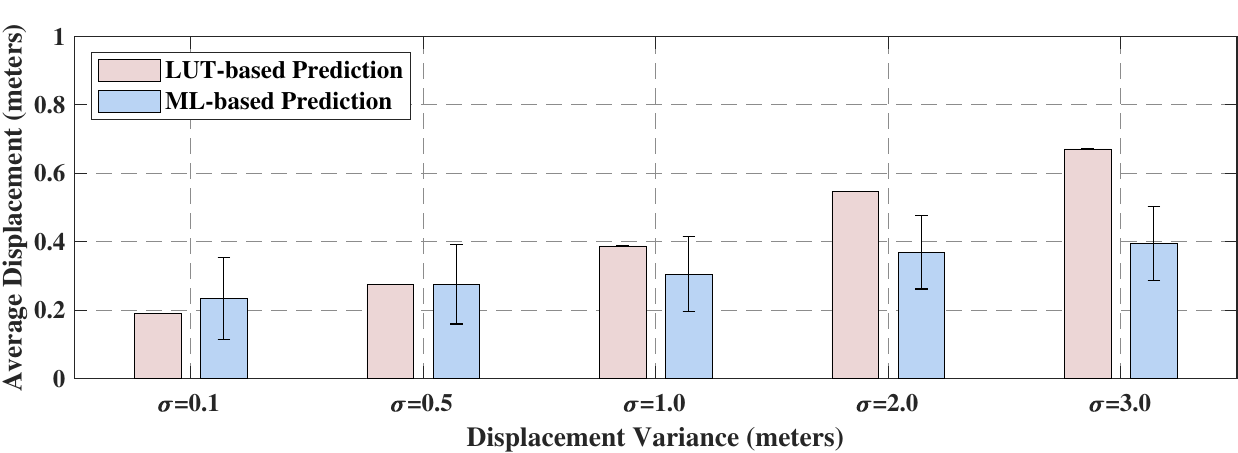}\label{fig:dist_plot_L2R}}
	\subfigure[Right-to-Left Dataset]{\centering \includegraphics[width=0.9\columnwidth]{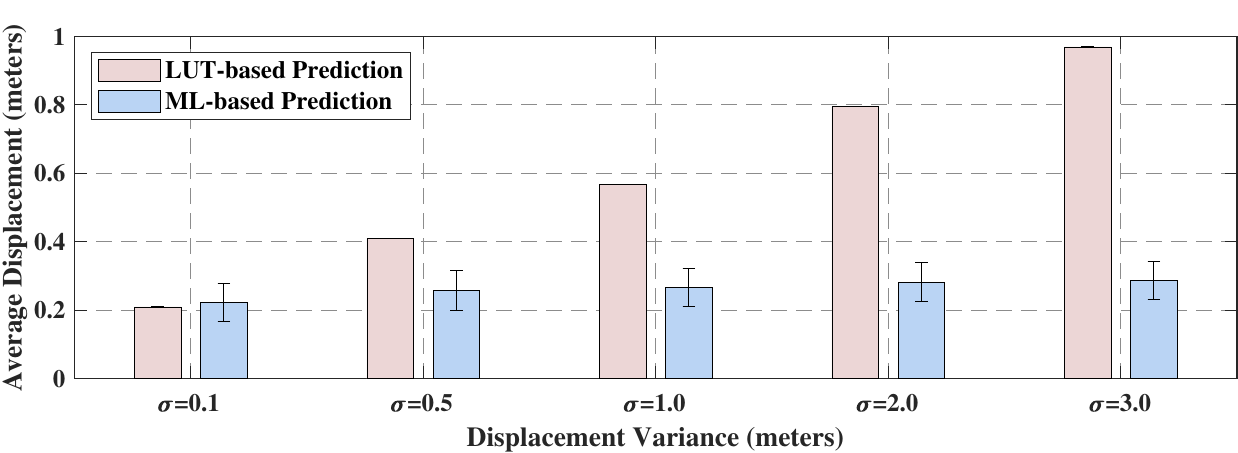}\label{fig:dist_plot_R2L}}
	
	\caption{De-noising performance comparison of the proposed solution for both datasets. }
	\label{fig:metric_comp}
\end{figure}

\textbf{Can additional sensory data help in user position de-noising?}
In this sub-section, we present the performance of our proposed vision-aided position de-noising solution on a real-world multi-modal dataset. As highlighted in Section~\ref{sec:dev_dataset}, errors with different variances are introduced to the ground-truth position data as part of the post-processing step. In Fig.~\ref{fig:dist_plot_L2R} and Fig.~\ref{fig:dist_plot_R2L}, we show the average displacement in meters for the ground-truth data and the predicted data for both the left-to-right and right-to-left datasets. The haversine distance between the ground-truth mean and predicted positions is computed for all the samples (test dataset) in each grid. The average across the $100$ grids is presented in both figures for all the error variances. It is observed that both LUT-based and deep learning-based approaches achieve comparable de-noising performance for the cases with lower error variance of $0.1$ and $0.5$ meters. The capability of the deep learning-based approach is highlighted for the cases with higher error variances. In Fig~\ref{fig:gmap_plot}, we present three images with the ground-truth, noisy ($\sigma=0.5$), and deep learning-based predicted positions overlaid on the Google Earth images. The results highlight the proposed solution's efficacy in predicting the user's de-noised position utilizing visual and wireless data.

\section{Conclusion}\label{sec:conc}

This paper introduced a novel, multi-modal vision-aided approach to significantly reduce GPS error margins, crucial for the development of smart cities and autonomous vehicle technologies. Our solution diverges from traditional methods by integrating machine learning with mmWave/THz wireless and visual data from a 5G and beyond communication system, offering a promising path to sub-meter GPS accuracy. Through site-specific error characterization and a comprehensive grid-based analysis, we demonstrated the effectiveness of our proposed solution in real-world scenarios. The findings highlight the potential of combining visual and wireless data to enhance GPS positioning, marking a significant step forward in achieving high-precision location services for future applications. 


\end{document}